\documentclass[12pt]{article}
\usepackage[utf8]{inputenc}
\usepackage{amssymb}
\usepackage{amsfonts}
\usepackage{amsmath}
\usepackage{graphicx}
\usepackage{multicol}
\usepackage{multirow}
\usepackage{lineno}
\usepackage{makecell}
\usepackage{natbib}
\usepackage{caption}
\usepackage{layouts}
\usepackage{algorithm}
\usepackage{algpseudocode}
\usepackage{url}
\setcitestyle{authoryear}
\usepackage{setspace}
\usepackage[vmargin=1in,hmargin=1in]{geometry}

\title{\normalsize \textbf{ 
Robust Privacy-Preserving Models for Cluster-Level Confounding: Recognizing Disparities in Access to Transplantation}}

\begin{document}

\date{}

\label{firstpage}


\maketitle

\vspace{-50pt}

\begin{center}
    Nicholas Hartman and
Kevin He$^*$ \\
Department of Biostatistics, University of Michigan, Ann Arbor, MI, U.S.A. \\
Kidney Epidemiology and Cost Center, University of Michigan, Ann Arbor, MI, U.S.A. \\
$^*${\it email:} kevinhe@umich.edu
\end{center}

\begin{abstract}
In applications where the study data are collected within cluster units (e.g., patients within transplant centers), it is often of interest to estimate and perform inference on the treatment effects of the cluster units. However, it is well-established that cluster-level confounding variables can bias these assessments, and many of these confounding factors may be unobservable. In healthcare settings, data sharing restrictions often make it impossible to directly fit conventional risk-adjustment models on patient-level data, and existing privacy-preserving approaches cannot adequately adjust for both observed and unobserved cluster-level confounding factors. In this paper, we propose a privacy-preserving model for cluster-level confounding that only depends on publicly-available summary statistics, can be fit using a single optimization routine, and is robust to outlying cluster unit effects. In addition, we develop a Pseudo-Bayesian inference procedure that accounts for the estimated cluster-level confounding effects and corrects  for the impact of unobservable factors. Simulations show that our estimates are robust and accurate, and the proposed inference approach has better Frequentist properties than existing methods. Motivated by efforts to improve equity in transplant care, we apply these methods to evaluate transplant centers while adjusting for observed geographic disparities in donor organ availability and unobservable confounders. 
\end{abstract}


\newpage

\section{Introduction}
\label{sec:intro}

Clustered data often arise in applications where the observations are collected within common units, such as institutions or geographic regions. In the study of organ failure, for example, patients that are treated by the same transplant center have outcomes which are more similar to each other than if they had been treated by different centers. In many instances, the treatment effects of these cluster units (e.g., transplant centers) are of interest, and the goal is to assess a particular aspect of the relationship between the cluster unit and the outcome variable. However, in observational studies, it is well-established that the inevitable influence of confounding variables can bias these treatment effect estimates \citep{Jones2011,He2013}, and many of these confounding variables may be defined at the level of the cluster unit instead of the observation unit. In addition, it is almost always the case that some of these confounding variables are unobservable. 

One important example of cluster-level confounding occurs in the treatment of End-Stage Renal Disease (ESRD), where the existence of geographic disparities in donor organ availability impacts patients’ access to life-saving transplants \citep{Wolfe}. In the United States (U.S.), the Centers for Medicare and Medicaid Services (CMS) encourages transplantation by using summary measures from statistical models to identify underperforming transplant centers and sanction those with low-quality care \citep{CMS}. Considering that these sanctions may include financial penalties or suspension, it is essential that the evaluations are based on accurate estimates of treatment quality, without the influence of confounding risk factors. One complication in these efforts is that there are substantial disparities in the availability of donor organs across U.S. geographic regions \citep{King,Hudgins}. Thus, centers that reside in regions with few available organs are disadvantaged in delivering transplants, even if they provide high-quality care. If these disparities are ignored, the most disadvantaged centers from certain regions may be unfairly penalized, exacerbating geographic inequities in access to transplantation. 

Under this setting, direct applications of conventional models for cluster-level confounding would require very large amounts of patient-level transplant and donation records from national registries, and for many researchers who perform routine evaluations, it is prohibitively expensive and time-consuming to repeatedly request and reanalyze this amount of protected data after each quarterly update of the registries. Therefore, many instead opt to use pre-calculated and publicly-available summary statistics to evaluate U.S. transplant centers \citep{SRTR_psr}, despite the fact that these statistics are typically not adjusted for important cluster-level confounding risk factors. To overcome the limitations of this approach, one may alternatively consider implementing privacy-preserving versions of conventional risk-adjustment models, which would not involve the direct use of the registry data. However, most existing privacy-preserving models rely on decentralized optimization approaches that would require the individual centers to coordinate a large number of repeated calculations and communications, which is widely recognized as an unrealistic scenario \citep{Jochems,Duan}. 

Other existing federated learning approaches depend on restrictive exchangeability conditions \citep{Han}, which assume that all confounding factors are observed and thus are not applicable to our motivating healthcare application where many confounding factors are potentially unobservable. For example, the COVID-19 pandemic has substantially impacted the U.S. transplantation network, but it is difficult to measure and quantify all of the confounding factors that have resulted from this event  \citep{Miller}. Furthermore, it has been shown that unobserved confounding can cause overdispersion in the test statistics that are commonly used to detect significant cluster unit treatment effects \citep{Spiegelhalter_Plot,Xia2020}. While individualized empirical null (EN) methods have been developed to estimate the severity of this overdispersion for each cluster unit, and correct the test statistics accordingly \citep{Hartman}, these methods are incapable of leveraging information from observed cluster-level confounding variables, resulting in a loss of power. In addition, EN methods rely on a Frequentist testing framework, which ignores variability in the confounding parameter estimates and increases the risk for false detection of low-quality care. Thus, new methods are needed that can estimate the effects of observed cluster-level confounding variables, incorporate these estimates into a valid inference procedure, and account for residual confounding variables that are unobservable.

To overcome these challenges, we derive a privacy-preserving model for cluster-level confounding estimation that only depends on public summary statistics and thus circumvents the practical obstacles in analyzing patient-level datasets. The required summary statistics are already available and familiar to policymakers, and the proposed estimates can be computed from a single optimization routine. By modeling functions of the summary statistics with asymptotic truncated normal densities, we develop estimators that are highly robust to outlying cluster unit effects. Finally, we propose a Pseudo-Bayesian inference method to detect underperforming providers while adjusting for the estimated cluster-level confounding effects and correcting for the additional impact of unobservable confounding variables on the posterior distributions. Simulations show that the proposed model accurately estimates the confounding effects, and the Pseudo-Bayesian evaluation method has a lower false-detection rate and a higher true-detection rate for significant cluster unit treatment effects, compared to conventional EN approaches. We apply these methods to evaluate the performances of U.S. transplant centers, with adjustments for observed geographic disparities in donor organ availability and other unobservable confounding factors.

\section{Methods}
\label{sec:methods}

\subsection{Background and Framework}
\label{sec:framework}

\subsubsection{Notation}
\label{sec:model}

Throughout this paper, we introduce our proposed methods in the context of applications where patient health outcomes are of interest and healthcare providers serve as the clustering units, though these methods may be applied in other settings. Let $i=1,\dots,I$ be the provider index and let $j=1,\dots,n_i$ be the patient index, where $n_i$ is the number of patient records within the $i^{th}$ provider. Assume that the outcome data $Y_{ij}$ are generated from a Generalized Linear Model with \begin{equation}
\label{eq:glm2} g(E[Y_{ij}|\theta_{ij}^*])=\theta_{ij}^*
   =\mu^*+\gamma^*_i+\boldsymbol{X}_{ij}^{\top} \boldsymbol{\beta}+\boldsymbol{W}_i^{\top}\boldsymbol{\nu}+\alpha_i,\end{equation}
\noindent where $g(\cdot)$ is the canonical link function, $\mu^*$ is the population norm, $\gamma^*_i$ is the treatment effect that reflects quality of care, $\boldsymbol{X}_{ij}$ is a vector of patient-level variables with effects $\boldsymbol{\beta}$, $\boldsymbol{W}_i$ is a vector of observed provider-level confounding variables with effects $\boldsymbol{\nu}$, and $\alpha_i$ is an unobserved random quantity that follows a $\textrm{N}(0,\sigma^2_{\alpha})$ distribution  (conditional on $\boldsymbol{X}_{i1},\dots,\boldsymbol{X}_{in_i}$ and $\boldsymbol{W}_i$). The assumption that $\alpha_i$ has a conditional mean of zero implies that $\alpha_i$ is independent of $\boldsymbol{X}_{i1},\dots,\boldsymbol{X}_{in_i}$ and $\boldsymbol{W}_i$ (we discuss in Section \ref{sec:cre} how this assumption can easily be relaxed so that our method can be applied to situations where $\alpha_i$ is correlated with the observed variables). Furthermore, define the function $b(\cdot)$, with $b'(\cdot)=g^{-1}(\cdot)$, and let $a(\psi)$ be a prespecified function of the nuisance parameter, $\psi$. We assume that $\boldsymbol{W}_i,i=1,\dots,I$ are independent and identically distributed and, without loss of generality, $E[\boldsymbol{W}_i]=\boldsymbol{0}$. In situations where the second condition does not hold, $\mu^*$ can be replaced with $\mu=\mu^*+E[\boldsymbol{W}_i^{\top}\boldsymbol{\nu}]$ and $\boldsymbol{W}_i$ can be replaced with a centered version. 

\subsubsection{Sources of Variation}
\label{sec:sources}

The terms in (\ref{eq:glm2}) are responsible for several sources of variation in $Y_{ij}$ across the providers:  \begin{enumerate}
\item[(i)] Provider treatment effects that are clinically meaningful and depart substantially from national norms ($\gamma_i^* $ values that are far from zero). 
\item[(ii)] Provider treatment effects that are are not clinically meaningful and are similar to national norms ($\gamma_i^*$ values that are very close to or exactly zero). 
\item[(iii)] Differences in the patient case-mixes of the providers (variation in $\boldsymbol{X}_{i1},\dots,\boldsymbol{X}_{in_i}$ across providers). For example, some transplant centers may treat more patients of certain blood types that are more difficult to transplant. 
\item[(iv)] Observed cluster-level confounding factors ($\boldsymbol{W}_i,i=1,\dots,I$), with effects $\boldsymbol{\nu}$,  which impact patient outcomes and are unrelated to quality of care. For example, $\boldsymbol{W}_i$ may represent the observed geographic disparities in donor organ availability that affect transplant care but are beyond the centers' control. 
\item[(v)] Unobservable cluster-level confounding ($\alpha_i, i=1,\dots,I$), which impacts patient outcomes and is unrelated to quality of care. For example, $\alpha_i$ may represent a combination of many factors that are difficult to measure, such as the complex impact of the COVID-19 pandemic on the U.S. transplantation network. 
\end{enumerate}
\noindent 
The objective is to isolate the variation from source (i) above and identify outlying providers with substantial deviations from the national norms of healthcare quality. Sources (i) and (ii) can be modeled by two different frameworks, which we describe further in Section \ref{sec:normal}.  

\subsubsection{Available Summary Data}
\label{sec:summary}

As discussed in Section \ref{sec:intro}, U.S. national transplant registries are maintained by a very limited number of entities, and for most policymakers, it is typically infeasible to frequently request this massive amount of protected patient-level data for routine monitoring of transplant center performance. Therefore, stakeholders within the transplantation community depend on center-level summary statistics from the Scientific Registry of Transplant Recipients (SRTR), a contractor of the U.S. Department of Health and Human Services that is responsible for using the national registry data to report on the quality of providers within the transplant system \citep{SRTR_psr}. These reports are publicly-available and updated quarterly.

For a certain patient outcome of interest such as transplantation, death, or graft failure, the SRTR reports the number of outcome occurrences observed within each U.S. transplant center for a given time period. In addition, the SRTR computes the number of outcome occurrences that are expected if the center were to provide care consistent with national norms. This expected number is generated from a risk-adjustment model that controls for differences in the centers' patient case-mixes. For example, some centers may treat many patients with pre-existing high-risk conditions, which could impact the observed patient outcomes despite being unrelated to the centers' quality of care. Conventional analyses of transplant center performance are based on indirect standardization where the observed and expected outcomes within the centers are compared to each other. 

Along with the observed and expected outcomes described above, one may obtain information on the centers' effective sizes, which are related to the variances of the observed outcomes. Other variables related to the  characteristics of the centers and their geographic regions, which could potentially serve as the $\boldsymbol{W}_i$, are also available from the public reports \citep{SRTR_psr}. Using the notation described in Section \ref{sec:model}, we define the available summary statistics formally as $O_i=\sum_{j=1}^{n_i} Y_{ij}$, $E_i=\sum_{j=1}^{n_i} b'(\theta^0_{ij})$, and $\tilde{n}_i=\sum_{j=1}^{n_i} b''(\theta^0_{ij})$, where  $\theta^0_{ij}=\mu^*+\boldsymbol{X}_{ij}^{\top} \boldsymbol{\beta}$, $O_i$ is the observed outcome, $E_i$ is the expected outcome, and $\tilde{n}_i$ is the center's effective size. In our motivating application, we study patients' access to transplantation across the U.S., so the main outcome of interest is the delivery of transplants to the centers' patient populations. 

It is important to note that the $E_i$ from the SRTR's reports is only adjusted for observable patient characteristics that are known to be clinically relevant. Thus, conventional evaluations based on the SRTR's reports account for source (iii) variation (Section \ref{sec:sources}) that is due to differences in patient case-mix, but they do not account for source (iv) or (v) variation caused by observed or unobserved cluster-level confounding effects. As described in Section \ref{sec:intro}, EN methods \citep{Hartman} have been proposed to account for both source (iii) and (iv) variation, but they cannot account for source (v) variation. These EN methods also require that $\alpha_i$ is independent of the observed variables and the estimated confounding effect parameters are precise enough to be treated as known quantities. We discuss these limitations further and address them in Sections \ref{sec:cre} and \ref{sec:Frequentist}.

\subsubsection{Fixed, Random, and Correlated Random Effects}
\label{sec:mods}

The risk-adjustment models used to calculate centers' expected outcomes can be constructed in several different ways. Conventional fixed effects (FE), random effects (RE), and correlated random effects (CRE) risk-adjustment models differ in how they specify the $\gamma_i^*$ term in (\ref{eq:glm2}), and these differences are directly related to the models' properties and limitations \citep{Kalbfleisch2013}. The FE model treats $\gamma_i^*,$ $i=1,\dots,I$ as fixed parameters and provides unbiased estimates of $\boldsymbol{\beta}$, but it is overspecified, since $\boldsymbol{W}_i$ is also measured at the provider-level and the data do not contain sufficient information to separate the confounding effects of $\boldsymbol{W}_i$ from $\gamma^*_i$. 

An alternative modeling strategy is to treat $\gamma_i^*$ as a random quantity and $\boldsymbol{W}_i$ as a fixed covariate. This RE model circumvents the overspecification issues of the FE model \citep{Wooldridge}, but it generally relies on the assumptions that $\gamma_i^*$ is independent of the fixed covariates and that $\gamma_i^*$ is generated from a common normal distribution for all $i$, which usually do not hold in provider profiling applications \citep{Kalbfleisch2013}. For example, the best-performing providers may attract certain types of patients, which violates the independence assumption, and there are almost always providers with outlying treatment effects, so it may be inappropriate to model the $\gamma_i^*$ with one common distribution. 

The CRE model extends the RE model to allow for a specific correlation structure between the random effects and the fixed covariates \citep{Neuhaus}. However, the coefficient estimates may suffer from a severe lack of precision if there are outlying providers with $\gamma_i^*$ that deviate from the assumed common normal distribution \citep{Finch,lmm}, as we show through simulation in Section \ref{sec:cre_sim}. This imprecision is especially problematic for conventional evaluations based on Frequentist testing approaches (Section \ref{sec:Frequentist}), since these methods assume that the model coefficients are very precisely estimated and can be treated as known constants. While robust CRE models have been proposed to mitigate this limitation \citep{Finch,robustlmm,rlme}, we show in Section \ref{sec:cre_sim} that the implementation of these methods is computationally infeasible for large-scale applications such as national transplant research. 

The most severe limitations of all these models are that data sharing restrictions in our motivating application prevent us from fitting them directly to patient-level data, and both observed and unobserved cluster-level confounders can bias the treatment effect estimates. In Section \ref{sec:estim}, we overcome these challenges by proposing a modeling approach which only relies on publicly-available summary statistics that are familiar to stakeholders, leverages the advantageous properties of the FE, RE, and CRE models, and seamlessly accounts for outlying provider effects. As a first step, we obtain center-level summary statistics from the SRTR's underspecified FE model, which ignores $\boldsymbol{W}_i$ and $\alpha_i$, but achieves unbiased and precise estimates of $\boldsymbol{\beta}$ without any distributional assumptions on $\gamma_i^*$. Then, using these naive test statistics, we derive a model to estimate $\boldsymbol{\nu}$ and allow for correlation between $\gamma_i^*$ and $\boldsymbol{X}_{i1},\dots,\boldsymbol{X}_{in_i}$ in the same way as the CRE model, while using asymptotic truncated normal densities to explicitly model outlying providers, correct for the impact of unobserved confounding, and provide stable estimation in the presence of extreme quality of care effects. Throughout the remainder of this paper, we refer to our proposed method as a Robust Privacy-Preserving Cluster-Level Confounding (RPP-CLC) model.  

\subsection{Estimation}
\label{sec:estim}

\subsubsection{Naive Z-Scores}
\label{sec:lin}

\noindent Our objective is to use the SRTR's summary-level data to derive a likelihood function involving the cluster-level confounding effect parameters of interest, $\boldsymbol{\nu}$ and $\sigma^2_{\alpha}$. As a first step, we construct naive standardized Z-scores from an FE score test, based on the misspecified null hypotheses $H_{0i}: \gamma_i=0, i=1,\dots,I$, with $\gamma_i=\gamma_i^*+\alpha_i+\boldsymbol{W}_i^{\top}\boldsymbol{\nu}$: \begin{equation}
        Z_{FE,i}=\frac{O_i-E_i}{\sqrt{a(\psi)\tilde{n}_i}}.
        \label{eq:Z_FE}
\end{equation}

\noindent We refer to these Z-scores as naive because they are only adjusted for observed patient-level factors, as discussed in Section \ref{sec:summary}. However, as we will see, these statistics can serve as useful intermediate data for estimating the cluster-level confounding effect parameters. 

\subsubsection{Functional Form: Normal Distribution}
\label{sec:normal}

We now derive useful conditional moments of the naive Z-scores in (\ref{eq:Z_FE}) and show that they involve $\boldsymbol{W}_i$, $\boldsymbol{\nu}$, and $\sigma^2_{\alpha}$.  First consider the case where $Y_{ij}|\theta^*_{ij}$ follows a normal distribution; we show in Web Appendix A that \begin{equation}
E[Z_{FE,i}|\gamma_i^*=0]=\sqrt{\frac{n_i}{\sigma^2_{\varepsilon}}}\boldsymbol{W}_i^{\top} \boldsymbol{\nu} \textrm{ \hspace{14pt}    and   \hspace{14pt}   } Var[Z_{FE,i}|\gamma_i^*=0]=1+\varphi n_i,
\label{eq:normal}
\end{equation}

\noindent where $\varphi=\sigma^2_{\alpha}/\sigma^2_{\varepsilon}$ and $\sigma^2_{\varepsilon}=a(\psi)$ is the nuisance parameter from (\ref{eq:glm2}). Therefore, for providers with $\gamma_i^*=0$, the naive Z-scores follow a provider-level linear model with coefficient vector $\boldsymbol{\nu}$. In Section \ref{sec:with_outliers}, we use this result to construct a likelihood function for $\boldsymbol{\nu}$ and $\sigma^2_{\alpha}$. 

The formulas in (\ref{eq:normal}) are written under the condition that $\gamma_i^*=0$, which implies that the center provides care exactly equal to the national expectations. Alternatively, these formulas also hold if $\gamma_i^*$ is a random quantity with $E[\gamma_i^*|\boldsymbol{W}_i,\boldsymbol{X}_{i1},\dots,\boldsymbol{X}_{in_i}]=0$. Under this model, the difference between $\gamma_i^*$ and zero is due to fluctuations in healthcare quality that are not clinically meaningful (i.e., source (ii) from Section \ref{sec:sources}).
If we define $Var[\gamma_i^*|\boldsymbol{W}_i,\boldsymbol{X}_{i1},\dots,\boldsymbol{X}_{in_i}]=\sigma^2_{\gamma^*}$ for these providers contributing to source (ii) variation, then $\varphi$ becomes $(\sigma^2_{\alpha}+\sigma^2_{\gamma^*})/\sigma^2_{\varepsilon}$ and all other terms in (\ref{eq:normal}) are unchanged. Throughout this paper, we refer to centers with $\gamma^*_i$ equal to zero (using fixed $\gamma_i^*$) or with a conditional mean of zero (using random $\gamma_i^*$) as ``null centers" or ``average centers" which contribute to source (ii) variation, and we refer to all other centers as ``outliers'' which contribute to source (i) variation. For simplicity, we introduce our methods under the framework where $\gamma_i^*$ is fixed. 

\subsubsection{Functional Form: Poisson Distribution}
\label{sec:poisson}

We now consider the setting in which $Y_{ij}|\theta^*_{ij}$ follows a Poisson distribution, which is the most widely assumed model for transplant outcomes in our motivating application \citep{SRTR_psr}. The canonical log-link function allows us to derive exact expressions for the null moments: \begin{equation}
E[Z_{FE,i}|\gamma_i^*=0]=\sqrt{\tilde{n}_i}(\exp(\boldsymbol{W}_i^{\top}\boldsymbol{\nu}+\sigma^2_{\alpha}/2)-1 ),
\label{eq:poisson}
\end{equation} \noindent \begin{equation}
Var[Z_{FE,i}|\gamma_i^*=0]=\exp(\boldsymbol{W}_i^{\top}\boldsymbol{\nu}+\sigma^2_{\alpha}/2)\left[1+\exp(\boldsymbol{W}_i^{\top}\boldsymbol{\nu}+\sigma^2_{\alpha}/2)(\exp(\sigma^2_{\alpha})-1)\tilde{n}_i\right],
\label{eq:poisson_var}
\end{equation}

\noindent where $\tilde{n}_i=\sum_{j=1}^{n_i} \exp(\theta^0_{ij})$ (Web Appendix B). These formulas can also be extended to the Quasi-Poisson model by introducing $a(\psi)=\psi$ as an overdispersion parameter in (\ref{eq:Z_FE}).

\subsubsection{Approximate Functional Form: Exponential Family}
\label{sec:family}

For any outcome distribution in the exponential family, we can approximate the null mean and variance functions of $Z_{FE,i}$ using first-order Taylor series expansions of $b'(\theta^*_{ij})$ and $b''(\theta^*_{ij})$ around $\theta^0_{ij}$:
\begin{equation}
 E[Z_{FE,i}|\gamma^*_i=0]\approx \sqrt{\frac{\tilde{n}_i}{a(\psi)}}\boldsymbol{W}_i^{\top}\boldsymbol{\nu},
    \label{eq:lin}
\end{equation}
\noindent \begin{equation}Var[Z_{FE,i}|\gamma^*_i=0] \approx 1+\frac{\sum_{j=1}^{n_i}b'''(\theta^0_{ij})}{\tilde{n}_i} \boldsymbol{W}_i^{\top} \boldsymbol{\nu}+\varphi \tilde{n}_i,\end{equation}

\noindent where $\varphi=\sigma^2_{\alpha}/a(\psi)$ and $\tilde{n}_i=\sum_{j=1}^{n_i} b''(\theta^0_{ij})$ (Web Appendix C). 

In (\ref{eq:glm2}), we assumed that the distribution of $\alpha_i$, conditional on the observed covariates, is $\textrm{N}(0,\sigma^2_{\alpha})$.  However, the formulas in (\ref{eq:normal}) and (\ref{eq:lin}) also hold under the milder moment conditions that $E[\alpha_i|\boldsymbol{X}_{i1},\dots,\boldsymbol{X}_{in_i},\boldsymbol{W}_i]=0$ and $Var[\alpha_i|\boldsymbol{X}_{i1},\dots,\boldsymbol{X}_{in_i},\boldsymbol{W}_i]=\sigma^2_{\alpha}$. Thus, for applications in which the normal distributional assumption is suspect, it is still possible to accurately estimate $\boldsymbol{\nu}$ and $\sigma^2_{\alpha}$ using our proposed methodology. On the other hand, the normal distribution is widely used to model unobserved random variables that are outside of the providers' control \citep{Jones2011,Xia2020,Kalbfleisch2018}, and we argue that this is a reasonable assumption in many settings. 

\subsubsection{Model Estimation}
\label{sec:with_outliers}

If all centers were null, then the mean and variance formulas in Sections \ref{sec:normal}-\ref{sec:family} would hold for all providers, and the likelihood function for $\boldsymbol{\nu}$ and $\sigma^2_{\alpha}$ could simply be written as a product of normal densities. With formulas (\ref{eq:normal}) and (\ref{eq:lin}), the maximum likelihood estimators (MLEs) in this case would also be ordinary least squares estimators. The problem with this approach is that there are almost always outlying providers for which the mean and variance formulas do not hold, causing these ``normal MLEs" to be biased. Furthermore, we do not know exactly which providers are outliers with true deviations from national expectations, as this is the main goal of our inference procedure. 

To overcome this challenge, we propose a robust version of the privacy-preserving model that recognizes outlying providers through EN estimation strategies \citep{Efron,efron2007size,Xia2020}. First, we specify an interval $[A_{i},B_{i}]$ for each provider and assume that the $Z_{FE,i}$ scores for an outlying provider fall outside of this interval with probability one. Then, we use asymptotic truncated normal densities to model the Z-scores that fall within the null intervals. Extending the EN likelihood function from \citet{Hartman}, we have \begin{equation}
L(\boldsymbol{\nu},\sigma^2_{\alpha},\pi_0)=\prod_{i \in S_0} \pi_0 \phi_i(Z_{FE,i}; \boldsymbol{\nu},\sigma^2_{\alpha}) \prod_{i \notin S_0} \{1-\pi_0Q_i(\boldsymbol{\nu},\sigma^2_{\alpha})\},
\label{eq:likelihood}
\end{equation}

\noindent where $S_0=\{i: Z_{FE,i} \in [A_i, B_i]\}$ with cardinality $I_0$, $\phi_i$ is the normal density with mean and variance as defined in Sections \ref{sec:normal}-\ref{sec:family}, $Q_i=\int_{A_i}^{B_i} \phi_i(z) dz$, and $\pi_0$ is the null proportion. We maximize (\ref{eq:likelihood}) with respect to $\boldsymbol{\nu}$, $\varphi$, and $\pi_0$ using numerical optimization. In our algorithm, we define initial values for the parameters by leveraging the approximate mean function in (\ref{eq:lin}) and fitting a robust linear regression model. Details are provided in Web Appendix D. 

\subsubsection{Connection With the CRE Model and Endogeneity}
\label{sec:cre}

In this section, we consider the scenario in which $Y_{ij}$ is generated from an underlying ``between-within" model \citep{Wooldridge,Kalbfleisch2013}, where $\boldsymbol{X}_{ij}$ is decomposed into between- and within-provider components, and we show that it has a  connection with our RPP-CLC model. For simplicity, we omit $\boldsymbol{W}_i$ and $\alpha_i$ from this discussion. The CRE model allows $\gamma_i^*$ to be correlated with $\boldsymbol{X}_{ij}$ through $\boldsymbol{\bar{X}}_i$ by assuming that $\gamma_i^*=\boldsymbol{\bar{X}}_i^{\top}\boldsymbol{\xi}+\tau_i$, where $\boldsymbol{\bar{X}}_i=\sum_{j=1}^{n_i} \boldsymbol{X}_{ij}/n_i$ and $\tau_i|\boldsymbol{X}_{i1},\dots,\boldsymbol{X}_{in_i} \sim N(0,\sigma^2_{\tau})$ \citep{Neuhaus,Wooldridge}. Thus, \begin{equation}
\theta^*_{ij}=\mu^*+\tau_i+\boldsymbol{X}_{ij}^{\top}\boldsymbol{\beta}+\boldsymbol{\bar{X}}_i^{\top}\boldsymbol{\xi}, \label{eq:cre} \end{equation}

\noindent which can be reparameterized in terms of the between- and within-provider effects of $\boldsymbol{X}_{ij}$. 

If we treat $\boldsymbol{\bar{X}}_i$ as an observed provider-level confounding variable, then we may proceed with our proposed estimation procedure by first obtaining summary statistics from an underspecified FE model that ignores $\boldsymbol{\bar{X}}_i$. It is well-known that $\boldsymbol{\beta}$ is unbiasedly estimated in this FE model \citep{Neuhaus,Wooldridge}. Then, under the CRE model assumptions, we show in Web Appendix E that the functional form of our proposed RPP-CLC model can be expressed as $E[Z_{FE,i}|\boldsymbol{\bar{X}}_i] \approx \sqrt{\frac{\tilde{n}_i}{a(\psi)}} \boldsymbol{\bar{X}}_i^{\top}\boldsymbol{\xi}$. The exact versions of the RPP-CLC model for the Normal and Poisson outcome distributions can be derived similarly. Therefore, our RPP-CLC approach is an alternative method for fitting CRE models, with the additional advantages that it only requires provider-level summary statistics and it incorporates EN estimation methods to robustly model outlying providers. In Section \ref{sec:cre_sim}, we compare our RPP-CLC model with robust versions of the CRE model, and we find that our proposed model has advantages in terms of estimation stability and computational efficiency, even if the full patient-level data are available.

We originally assumed in Model (\ref{eq:glm2}) that $E[\alpha_i|\boldsymbol{X}_{i1},\dots,\boldsymbol{X}_{in_i},\boldsymbol{W}_i]=0$, which implies that $\alpha_i$ is independent of the observed covariates. However, by adopting aspects of the CRE model, this assumption can be relaxed to allow for correlation between $\alpha_i$ and the observed covariates. We may let $\alpha_i=\boldsymbol{\bar{X}}_i^{\top}\boldsymbol{\xi}+\boldsymbol{W}_i^{\top}\boldsymbol{\zeta}+\tau_i$, where $\tau_i|\boldsymbol{X}_{i1},\dots,\boldsymbol{X}_{in_i}, \boldsymbol{W}_i \sim \textrm{N}(0,\sigma^2_{\tau})$, such that $\theta^*_{ij}=\mu^*+\gamma_i^*+\tau_i+\boldsymbol{X}_{ij}^{\top} \boldsymbol{\beta}+\boldsymbol{\bar{X}}_i^{\top}\boldsymbol{\xi}+\boldsymbol{W}_i^{\top}(\boldsymbol{\nu}+\boldsymbol{\zeta}),$ and $\tau_i$ serves as the unobserved quantity, satisfying all required conditions. This result is a distinction from existing EN methods, which strictly require the independence assumption for $\alpha_i$ \citep{Hartman}. 

\subsection{Inference}

\subsubsection{Frequentist Approach}
\label{sec:Frequentist}

After obtaining estimates of $\boldsymbol{\nu}$ and $\sigma^2_{\alpha}$, our objective is to incorporate them into a valid inference procedure to identify outlying providers while adjusting for observed and unobserved cluster-level confounding factors. Existing approaches that account for patient-level confounding rely on a Frequentist hypothesis testing framework, where it is assumed that $\boldsymbol{\beta}$ can be estimated very precisely and treated as a known quantity \citep{Spiegelhalter,Hartman}. The justification for this assumption is that $\widehat{\boldsymbol{\beta}}$ is typically derived from millions of patient records included in the national registries. 

In contrast, the precision of $\widehat{\boldsymbol{\nu}}$ and $\widehat{\sigma}^2_{\alpha}$ increases with the number of unique providers instead of the number of patient records. Thus, even with massive patient-level datasets, the Frequentist testing approach is only appropriate for handling cluster-level confounding when a very large number of providers is under evaluation. If one were to proceed with this approach, a corrected version of the Z-scores, $Z_{FE,i}^*$, could be computed as \begin{equation} Z^*_{FE,i}=\frac{Z_{FE,i}-\widehat{E}[Z_{FE,i}|\gamma_i^*=0]}{\sqrt{\widehat{Var}[Z_{FE,i}|\gamma^*_i=0]}},
\label{eq:Z_FE2}
\end{equation}
where $\widehat{E}[Z_{FE,i}|\gamma_i^*=0]$ and $\widehat{Var}[Z_{FE,i}|\gamma_i^*=0]$ are obtained by plugging $\widehat{\boldsymbol{\nu}}$ and $\widehat{\sigma}^2_{\alpha}$ into the conditional moments from Sections \ref{sec:normal}-\ref{sec:family}. One may test whether a provider's care deviates from the national norm by comparing $Z_{FE,i}^*$ to a quantile of the N(0,1) distribution.    

\subsubsection{Pseudo-Bayesian Approach for Poisson Outcomes}
\label{sec:Bayes}

In many applications, the number of providers under evaluation is modest, and there may be non-negligible variability in $\widehat{\boldsymbol{\nu}}$ and $\widehat{\sigma}^2_{\alpha}$ that should be accounted for. For example, in our motivating application, there are only 256 U.S. kidney transplant programs, and Frequentist testing methods are unsuitable. Furthermore, clinicians prefer to interpret quality of care using ratios of observed and expected outcomes, but it is unclear how to account for the impact of cluster-level confounding without first converting these ratios to the Z-score scale, which is substantially less clinically meaningful. Very small centers can also have unstable and highly extreme ratio values, further complicating the interpretations \citep{SRTR_Bayes}. To avoid the inappropriate use of Frequentist testing methods under these scenarios, we develop a Pseudo-Bayesian inference procedure that accounts for the uncertainty in the cluster-level confounding parameter estimates and can be interpreted on the measure ratio scale. We observe empirically that the issue of imprecision is much less severe for $\widehat{\sigma}^2_{\alpha}$ compared to $\widehat{\boldsymbol{\nu}}$, so to substantially simplify our solution, we focus on the uncertainty in $\widehat{\boldsymbol{\nu}}$ and treat $\sigma^2_{\alpha}$ as known. Here, we use the term ``Pseudo-Bayesian'' to emphasize that our posterior distributions are approximated based on this approach.

Under the Pseudo-Bayesian framework, $\boldsymbol{\nu}$ is a random quantity, and we propose a multivariate normal prior distribution for $\boldsymbol{\nu}$, $\textrm{MVN}(\boldsymbol{0},\boldsymbol{\Sigma_{\textrm{prior}}})$. Asymptotically, $\widehat{\boldsymbol{\nu}} \sim \textrm{MVN}(\boldsymbol{\nu},\boldsymbol{\Sigma_{\widehat{\nu}}})$, and we show in Web Appendix F that $\boldsymbol{\Sigma_{\widehat{\nu}}}$ can be precisely estimated by
$$\boldsymbol{\widehat{\Sigma}_{\widehat{\nu}}}=(\boldsymbol{W_{\textrm{null}}}^{\top}\boldsymbol{W_{\textrm{null}}})^{-1}\boldsymbol{W_{\textrm{null}}}^{\top}\boldsymbol{\widehat{\Omega} W_{\textrm{null}}}(\boldsymbol{W_{\textrm{null}}}^{\top}\boldsymbol{W_{\textrm{null}}})^{-1},$$
\noindent where $\boldsymbol{W_{\textrm{null}}}$ is an $I_0$ by $P$ matrix of the $P$ confounding variables (among null providers), $\boldsymbol{\widehat{\Omega}}_{ii}=1+\boldsymbol{W}_i^{\top} \widehat{\boldsymbol{\nu}}+\widehat{\varphi} \tilde{n}_i$, and $\boldsymbol{\widehat{\Omega}}_{i \ne j}=0$. We find that $\boldsymbol{\widehat{\Omega}}_{ii}$ is usually dominated by $\tilde{n}_i$, and the variability in $\widehat{\boldsymbol{\nu}}$ has negligible impact on $\boldsymbol{\Sigma_{\widehat{\nu}}}$. Thus, we treat $\boldsymbol{\Sigma_{\widehat{\nu}}}$ as a known nuisance parameter, though fully-Bayesian methods may be applied. Asymptotically, the posterior distribution of $\boldsymbol{\nu}$ is $ \textrm{MVN}(\boldsymbol{m_{\textrm{post}}},\boldsymbol{\Sigma_{\textrm{post}}})$, where $\boldsymbol{m_{\textrm{post}}}=\boldsymbol{\Sigma_{\textrm{prior}}}\left(\boldsymbol{\Sigma_{\textrm{prior}}}+\boldsymbol{\Sigma_{\widehat{\nu}}}\right)^{-1}\widehat{\boldsymbol{\nu}}$ and $\boldsymbol{\Sigma_{\textrm{post}}}=\left( \boldsymbol{\Sigma}^{-1}_{\textrm{prior}}+\boldsymbol{\Sigma}^{-1}_{\widehat{\boldsymbol{\nu}}} \right)^{-1}$. We then incorporate the unobserved quantity, $\alpha_i$, by deriving an approximate $\textrm{Lognormal}(\boldsymbol{W}_i^{\top}\boldsymbol{m_{\textrm{post}}},\boldsymbol{W}_i^{\top} \boldsymbol{\Sigma_{\textrm{post}} W}_i+\sigma^2_{\alpha})$ posterior distribution for a useful random variable,
$\Lambda_i=\exp(\boldsymbol{W}_i^{\top}\boldsymbol{\nu}+\alpha_i)$. 

Let $R_i$ denote a random variable that corresponds to the true value of the naive healthcare quality measure ratio, which is not adjusted for $\boldsymbol{W}_i$ or $\alpha_i$. For measures of access to transplantation, the SRTR uses a Gamma-Poisson prior-likelihood framework \citep{Jones2011,SRTR_Bayes} to specify the posterior distribution of $R_i$ (denoted as $f_{R_i}(r_i)$) as Gamma$(O_i+2,E_i+2)$. We extend this framework to derive the following approximate posterior distribution for a corrected random variable of healthcare quality, $R_i^*$: \begin{equation}f_{R^*_i}(r^*_i)=\int f_{R_i^*|\Lambda_i}(r^*_i|\lambda_i)f_{\Lambda_i}(\lambda_i) d\lambda_i,\label{eq:Pfun}\end{equation}

\noindent where $f_{R_i^*|\Lambda_i}(r^*_i|\lambda_i)$ is the Gamma$(O_i+2,E_i\lambda_i+2)$ density function and $f_{\Lambda_i}(\lambda_i)$ is the approximate Lognormal posterior density function for $\Lambda_i$ defined above (Web Appendix  G). Outlying providers are flagged based on the posterior credible intervals. 

\section{Simulations}
\label{sec:sim}

\subsection{Estimation}
\label{sec:est_nu}

We assessed the accuracy of our estimation procedure through numerical evaluations. The outcome data were simulated from a Poisson model for 200 providers, each with $n_i=n=100$. A single observed cluster-level confounding variable, $W_i$, was generated from a $\textrm{N}(0,1)$ distribution, and the unobserved quantity, $\alpha_i$, was generated from a $\textrm{N}(0,0.1)$ distribution. The effect of $W_i$ was set as $\nu=0.25$. For some proportion of providers, we set $\gamma^*_i=0$. Then, for the remaining providers, we set $\gamma^*_i=c+0.5W_i$, where $c$ is a non-zero constant. We varied the proportion of providers with $\gamma^*_i \ne 0$ (i.e., the ``outlier proportion") and the magnitude of $c$ for the outlying providers. For each setting, we computed $Z_{FE,i}$ values from (\ref{eq:Z_FE}), and estimated $\nu$ and $\sigma_{\alpha}^2$ using both the normal MLE and the proposed RPP-CLC methods described in Section \ref{sec:with_outliers}. 

\begin{figure}
    \centering
    \includegraphics[width=\textwidth]{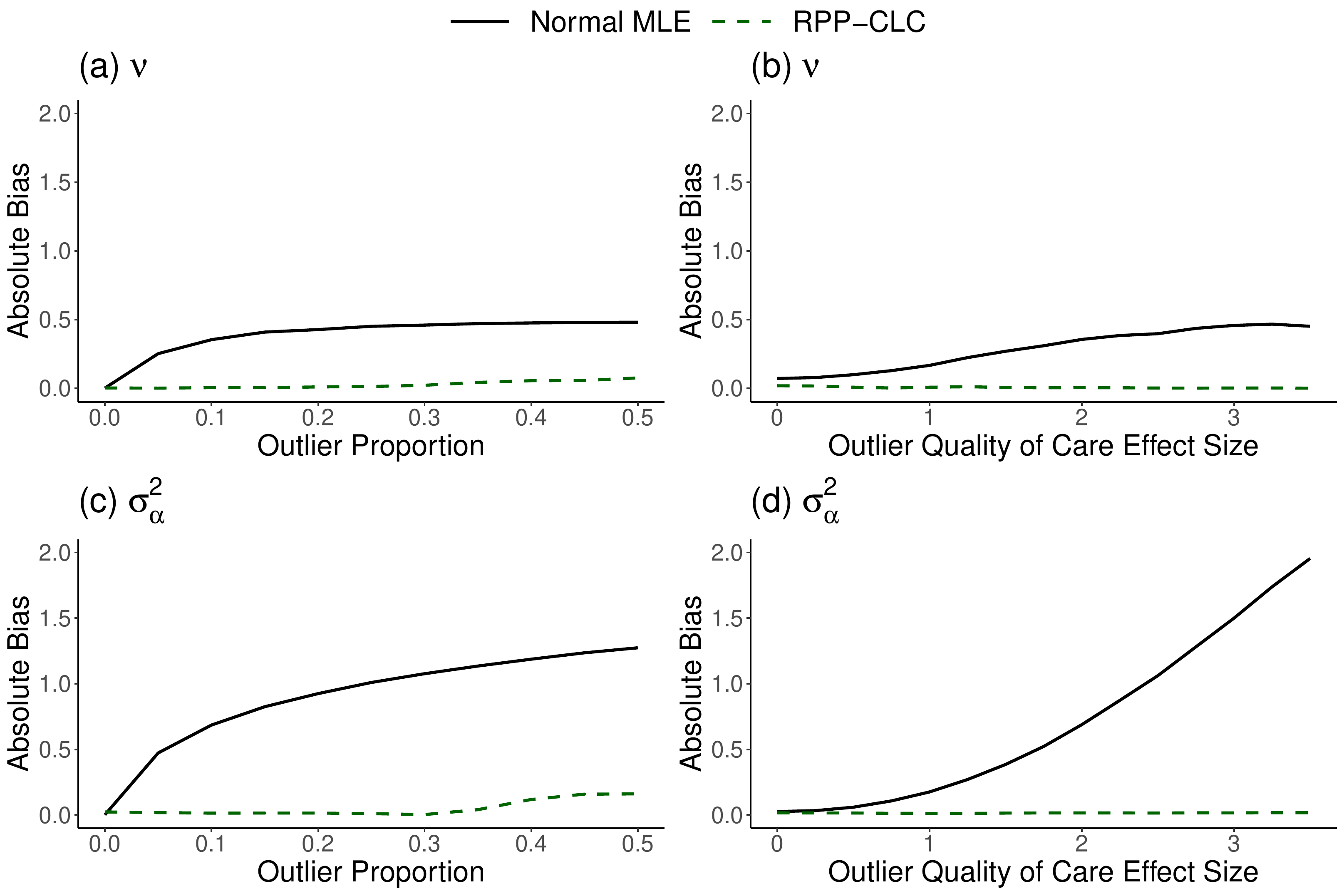}
    \caption{Absolute value of the bias in $\widehat{\nu}$ and $\widehat{\sigma}_{\alpha}^2$, estimated via the normal Maximum Likelihood Estimator (MLE) or the proposed Robust Privacy-Preserving model for Cluster-Level Confounding (RPP-CLC model). (a) and (c): the outlier quality of care effect size is fixed at two, and the outlier proportion is varied. (b) and (d): the outlier proportion is fixed at 0.1, and the outlier quality of care effect size is varied. Results are based on 1000 iterations.}
    \label{fig:Outlier_Sim}
\end{figure}

In Figure \ref{fig:Outlier_Sim}, we observed that both methods produced nearly-unbiased estimates of $\nu$ and $\sigma^2_{\alpha}$ when there were no outliers. However, as expected based on the arguments in Section \ref{sec:with_outliers}, the estimates from the normal MLE approach became much more biased than those from the RPP-CLC method as the outlier proportion and the outlier effect size increased. The RPP-CLC approach was highly robust to the outlying providers, and the estimates remained close to unbiased across all settings. 

\subsection{Comparisons With the CRE Model}
\label{sec:cre_sim}

In Section \ref{sec:cre}, we showed that our RPP-CLC model can estimate between-provider and within-provider covariate effects, as in the CRE model. We now compare the performances of these two methods in estimating $\boldsymbol{\xi}$, the coefficient for $\boldsymbol{\bar{X}}_i$ in (\ref{eq:cre}). We first simulated a patient-level covariate, $X_{ij}$, from a $\textrm{N}(m_X,0.25)$ distribution, where $m_X$ was drawn from a 
$\textrm{N}(-0.4,0.25)$ distribution. Then, we set $\mu^*=-6$, $\beta=1$, and $\xi=0.25$ in (\ref{eq:cre}). The random effects, $\gamma^*_i$, were generated as $\gamma^*_i=\xi \bar{X}_i+\tau_i$, where $\tau_i$ was simulated from a normal distribution contaminated by outliers. Here, we simulated $Y_{ij}$ from a normal distribution to facilitate later comparisons with robust CRE models, for which most public software are based on linear mixed models \citep{robustlmm,rlme}. 

\begin{figure}
    \centering
    \includegraphics[width=\textwidth]{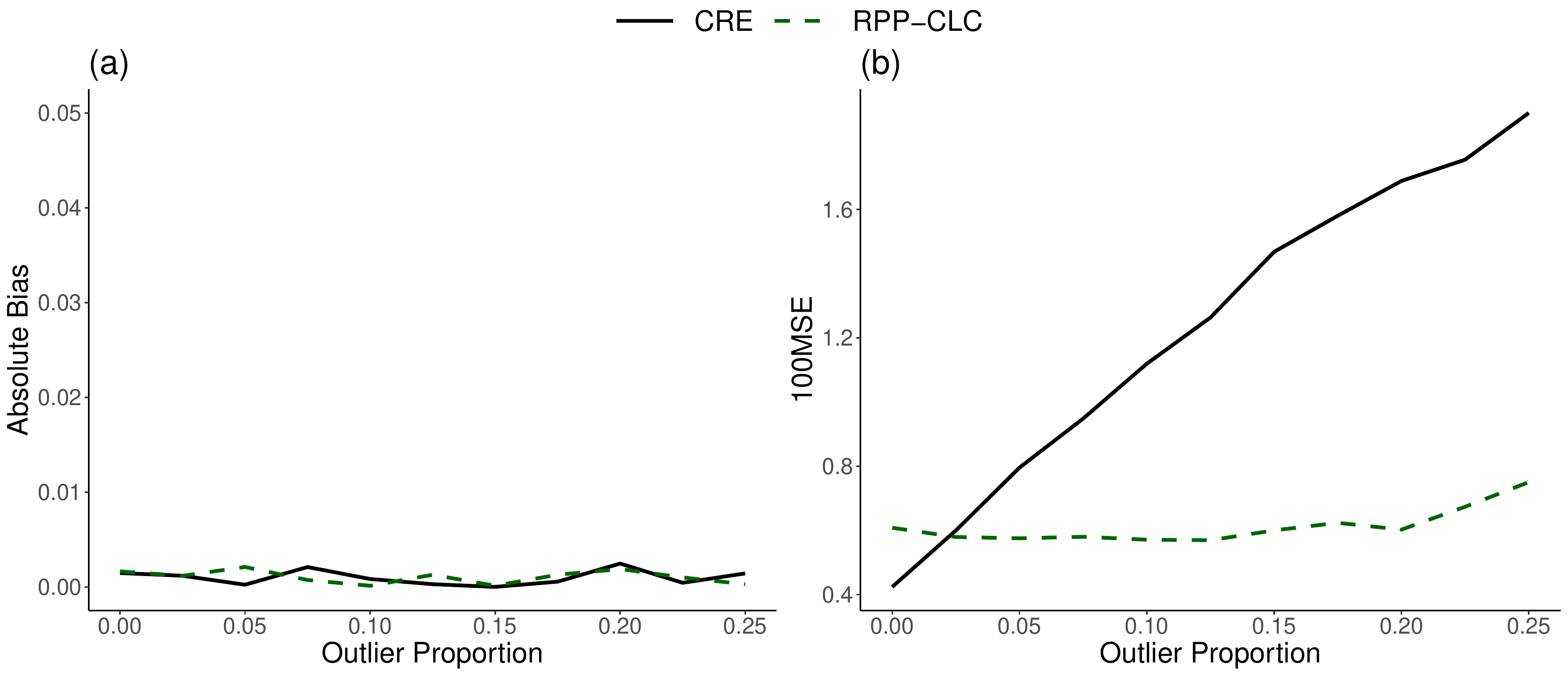}
    \caption{(a) Absolute bias and (b) 100 times the Mean Squared Error (100MSE) for the estimate of $\xi$ (the cluster-level confounding effect parameter in the underlying model), from either the Correlated Random Effects (CRE) model or the proposed Robust Privacy-Preserving model for Cluster-Level Confounding (RPP-CLC model). Results are based on 5000 iterations. The true confounding effect parameter value is $\xi=0.25$, and the proportion of providers with outlying cluster-level treatment effects is varied.}
    \label{fig:CRE_Sim}
\end{figure}

As shown in Figure \ref{fig:CRE_Sim}, the estimates from both the CRE model and the RPP-CLC model had low bias for every level of outlier contamination in the random effects. However, the CRE model estimates became much more unstable as outliers were introduced into the random effects distribution, and the MSE increased almost linearly with the outlier proportion. In contrast, the MSE for the RPP-CLC model remained constant, regardless of the outlier proportion (Figure \ref{fig:CRE_Sim}). Several authors have proposed robust versions of the CRE model to improve its precision \citep{Finch}. While these methods have theoretical validity, they are often computationally intensive for large-scale applications. Table \ref{tab:rt} compares the runtimes and memory usages of two popular robust versions of the CRE model \citep{robustlmm,rlme} and our proposed RPP-CLC model. All models were assessed in 64-bit R software \citep{R}. The robust CRE models were computationally expensive, and even with just 50,000 records, they exhausted the vector memory allocation in R (Table \ref{tab:rt}). 

In our motivating application, national transplant datasets contain millions of records, so we argue that robust patient-level CRE models are unsuitable for such large-scale applications, even if all patient-level data are available. Our RPP-CLC model is much more computationally efficient than the robust CRE models, and it is more stable in the presence of outliers, compared to the original CRE model. In addition, if patient-level data are restricted, our method can still be implemented using publicly-available summary statistics. 

\begin{table}
\centering
\caption{Runtime (and memory allocation) comparison among robust Correlated Random Effects (CRE) models and the proposed Robust Privacy-Preserving model for Cluster-Level Confounding (RPP-CLC model). 
Huberized CRE: robust scoring equations (\texttt{robustlmm} package), Rank CRE: rank-based CRE model (\texttt{rlme} package), RPP-CLC : proposed robust privacy-preserving model. Abbreviations: second (s), gigabyte (GB).}
\begin{tabular}{ccccc}
\hline
\textbf{Number of Records} & 10,000 & 50,000 & 100,000 \\
\hline
\textbf{Huberized CRE} & 280.2s (64.3GB) & Memory Exhausted  &  Memory Exhausted \\
\textbf{Rank CRE} & 5.61s (8.86GB) & Memory Exhausted & Memory Exhausted \\
\textbf{RPP-CLC} & 1.29s (1.36GB) & 2.09s (2.38GB) & 4.01s (2.33GB) \\
\hline
\end{tabular}
\label{tab:rt}
\end{table}

\subsection{Inference}
\label{sec:sflag}

Using a similar simulation structure as in Section \ref{sec:est_nu}, we explored the properties of our Frequentist and Pseudo-Bayesian flagging methods. We considered two different settings for the first provider in our simulated datasets. In the first setting, we defined a null provider with $\gamma_1^*=0$ and varied the magnitude of $W_1$. In the second setting, we let $W_1=1$, and we varied $\gamma^*_1$ so that the provider had increasingly low-quality care. For both the naive and adjusted Frequentist methods, we flagged the provider if the Z-score was more extreme than $\pm$1.96, which corresponds to a two-sided test at the 0.05 level. For the Pseudo-Bayesian methods, we flagged the provider if the $95\%$ credible interval did not contain one. 

\begin{figure}
    \centering
    \includegraphics[width=\textwidth]{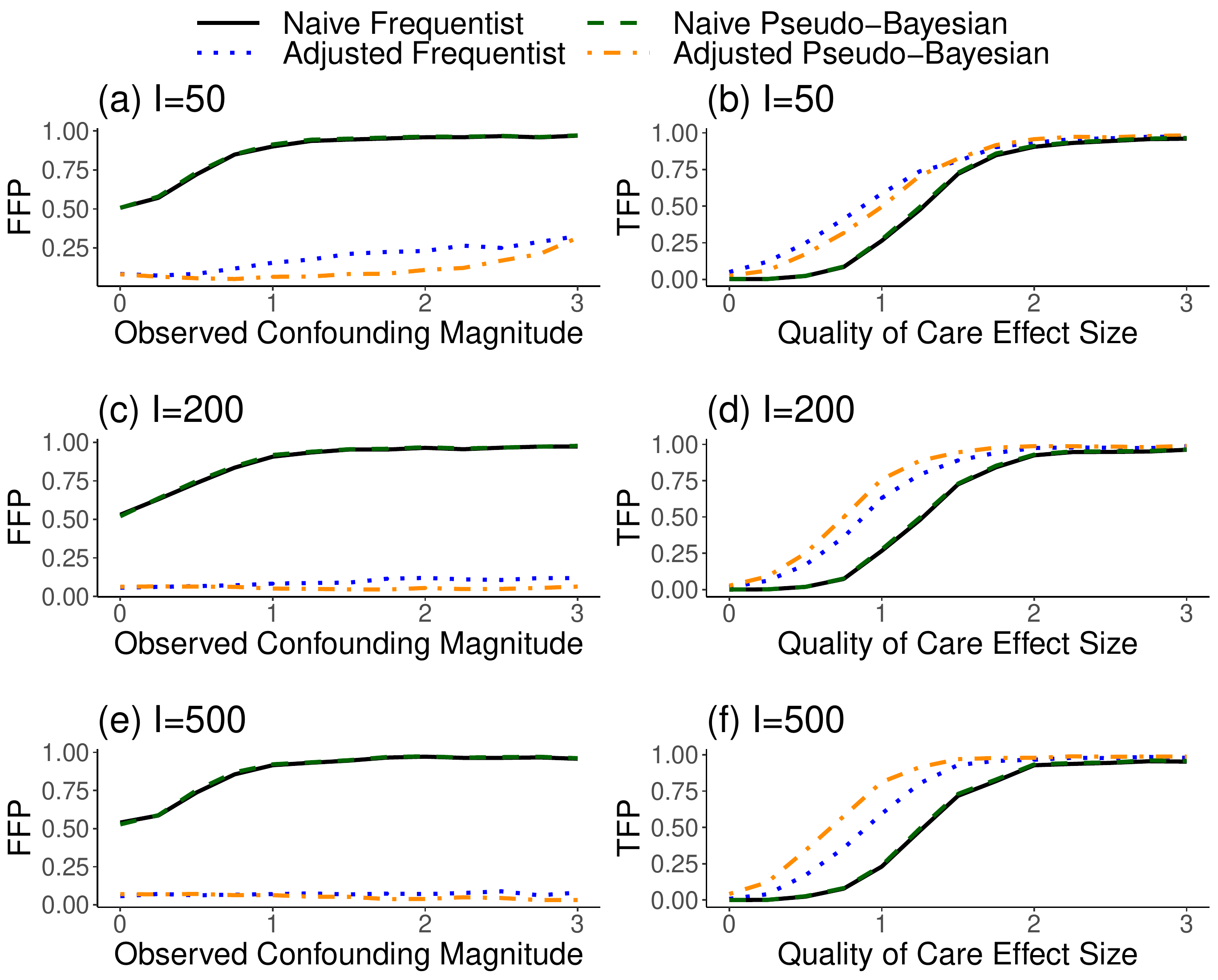}
    \caption{Probability of (a,c,e) falsely flagging a null provider or (b,d,f) correctly flagging a provider with low quality of care, for different levels of confounding or quality of care effect sizes. The Frequentist methods compare standardized Z-scores to an absolute threshold of 1.96. The Pseudo-Bayesian methods compare 95\% credible intervals to the null value of one. The naive methods ignore the observed and unobserved provider-level confounding, whereas the adjusted versions account for these variables. Results are based on 1000 iterations.  I: Number of Providers, FFP: False-Flagging Probability, TFP: True-Flagging Probability}
    \label{fig:Flag_Sim}
\end{figure}

Figure \ref{fig:Flag_Sim} shows that the ``naive" flagging approaches, which ignore $W_i$ and $\alpha_i$, had higher false-flagging probabilities (FFP) and lower true-flagging probabilities (TFP), compared to the adjusted versions. In general, the naive Frequentist and Pseudo-Bayesian methods produced very similar flagging results, but the adjusted Frequentist approach had a higher FFP relative to the adjusted Pseudo-Bayesian approach. The elevated FFP of the Frequentist approach was much more severe when the number of providers ($I$) was small. These findings are expected, since only the Pseudo-Bayesian approach accounts for uncertainty in $\widehat{\nu}$, and $\widehat{\nu}$ becomes less precise as $I$ decreases. In addition, the adjusted Pseudo-Bayesian approach had a higher TFP than the adjusted Frequentist approach for large $I$ (i.e., when there is less shrinkage towards the prior mean of zero). We suspect that this improved TFP is a reflection of the fact that the Pseudo-Bayesian approach uses information from the underlying model to incorporate $\alpha_i$ directly into the posterior distribution. 

\section{Kidney Transplant Center Evaluations}
\label{sec:data_analysis}

\subsection{Donor Organ Availability}

Organ Procurement Organizations (OPOs) are independent corporations that are responsible for collecting and allocating donor organs to transplant centers within a particular geographic region, referred to as the donation service area (DSA). It has been shown that OPOs vary in their abilities to recover organs at a high rate, and combined with geographic variation in mortality rates and attitudes towards donation, this causes substantial disparities in the availability of donor organs across DSAs \citep{King}. For transplant centers that reside in DSAs with severe shortages in donor organ supply, it is very challenging to meet the needs of their ESRD patients and provide transplants quickly, even if these centers deliver high-quality care. In this application, we study the impact of cluster-level confounding factors, such as geographic disparities in donor organ availability, on the Transplant Rate Ratio (TRR), which describes a transplant center's ability to deliver transplants efficiently.

Several authors have noted that the COVID-19 pandemic has accentuated these disparities in OPO performance and donor organ availability across the nation \citep{Miller, Hudgins,BloomWorks}. It has been argued that high-performing OPOs were more prepared for pandemic-related disruptions, and many have seamlessly adapted to COVID-19 mitigation efforts and drawn from an expanded donor pool, while under-performing OPOs have had more-challenging transitions and in some cases declining donation rates \citep{BloomWorks}. The impact of COVID-19 on the organ procurement system and access to transplantation is likely to be long-lasting, and because OPOs have full control of their own DSAs, statistical models of transplantation must consider donor organ availability at the DSA level as an important explanatory factor. 

Currently, the SRTR's transplant center quality metrics are only adjusted for patient-level risk factors, and the corresponding risk-adjustment models do not control for geographic disparities in donor organ availability. Web Figure 1 shows the distributions of the SRTR's posterior medians for the TRR measure, stratified by the regional supply of donor organs. These descriptive plots suggest that centers with larger supplies of donor organs tend to be evaluated much more favorably than those with fewer resources, and the SRTR's current evaluation system may in large part reflect the DSAs' donor organ supplies, as opposed to the transplant centers' efforts to expand access to transplantation. Using summary statistics from the SRTR's public reports and applying our proposed methods, we evaluate transplant center care while controlling for the availability of donor organs within the centers' DSAs. In this way, we aim to make the evaluations more fair for centers that have suffered from severe pandemic-related disruptions in their donor organ supplies, and to identify centers that have enough resources available to expand access to transplantation. 

\subsection{SRTR Data}
\label{sec:SRTR}

We collected center-level and DSA-level summary statistics from the January 2022 public releases of the SRTR's transplant program and OPO-specific reports \citep{SRTR_psr}. The observed and expected numbers of transplants for the centers were based on a two-year cohort of waitlisted patients from July 1, 2019 to  June 30, 2021. The DSA-level counts of donors and waitlisted patients, which we used to define our adjustment variable, were based on a one-year cohort from July 1, 2020 to June 30, 2021. Thus, the data were measured during the height of the pandemic. We defined the observed cluster-level adjustment variable, $W_i$, as the number of donors (meeting eligibility criteria) per patient on the transplant waiting list within the $i^{th}$ center's DSA. 

\subsection{Analysis Results}

\begin{figure}
    \centering
    \includegraphics[width=\textwidth]{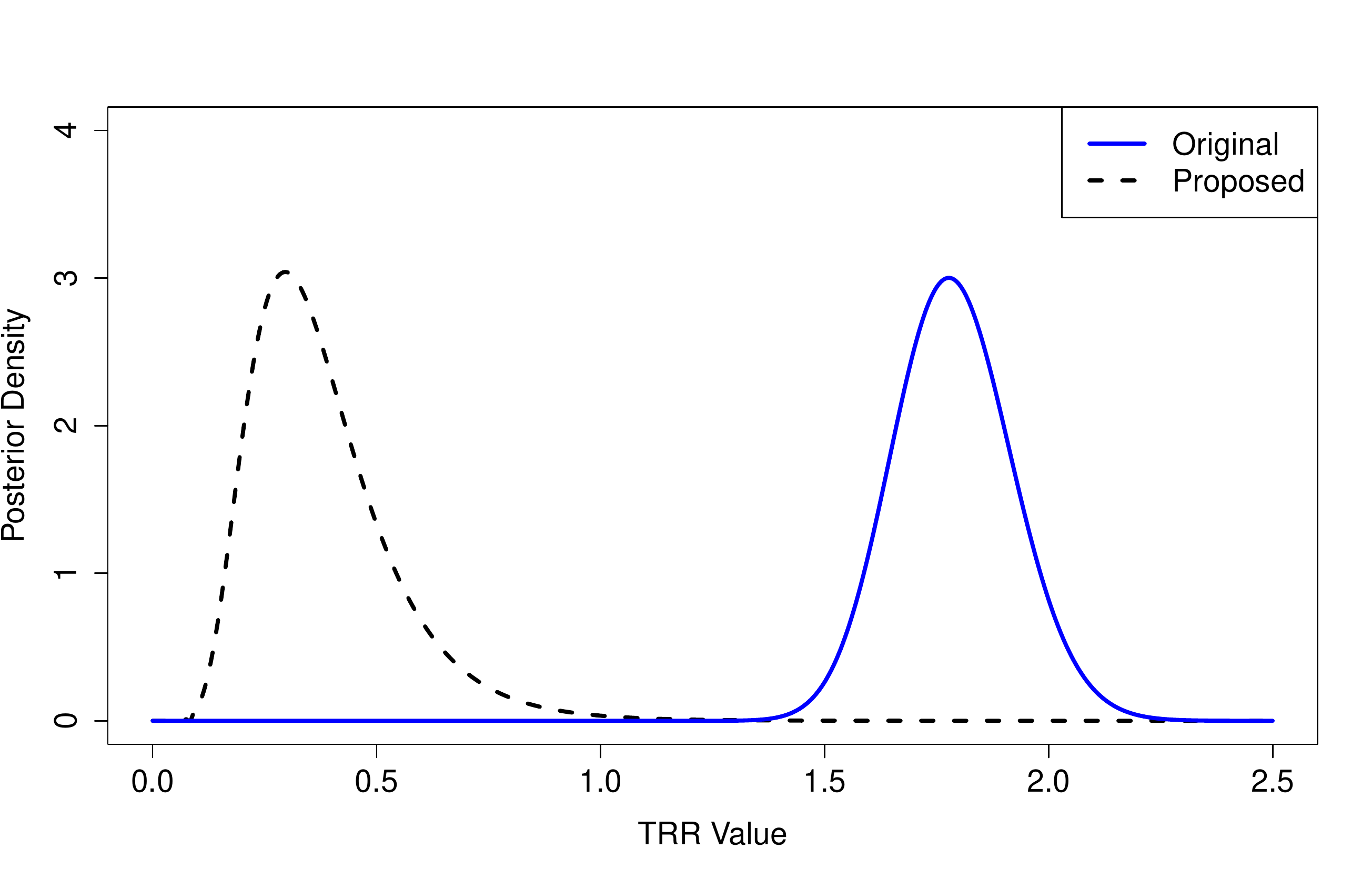}
    \caption{Original and proposed posterior distributions for the Transplant Rate Ratio (TRR) of one U.S. kidney transplant center. The proposed posterior distribution is adjusted for geographic disparities in donor organ availability and unobserved confounding factors, whereas the original posterior distribution is not adjusted for these factors. A TRR value of one indicates that the provider is performing consistently with the national norms, and lower TRR values indicate worse performance.}
    \label{fig:real_posterior_plot}
\end{figure}

We found that $W_i$ had a strong confounding effect on the TRR values. From our proposed estimation procedure, $\widehat{\nu}=4.47$ with a 95\% confidence interval of (3.40, 5.55). The example center in Figure \ref{fig:real_posterior_plot} resides in a DSA with very high donor organ availability, and while this center appeared to be performing transplants quickly according to the SRTR's original posterior distribution, $f_{R_i}(r_i)$, our proposed posterior distribution, $f_{R^*_i}(r^*_i)$, suggests that this center should be transplanting patients at a higher rate given the ample supply of donor organs in its DSA. Among the centers that were flagged as poor performers based on the credible intervals of $f_{R_i}(r_i)$, 79\% were reclassified as null centers based on the credible intervals of $f_{R^*_i}(r^*_i)$, which adjusts for the impacts of both $W_i$ and $\alpha_i$. From the results in Section \ref{sec:sflag}, we suspect that many of these centers were falsely flagged originally. We also identified seven new centers as poor performers after adjusting for the confounding effects. 

\begin{figure}
    \centering
    \includegraphics[width=\textwidth]{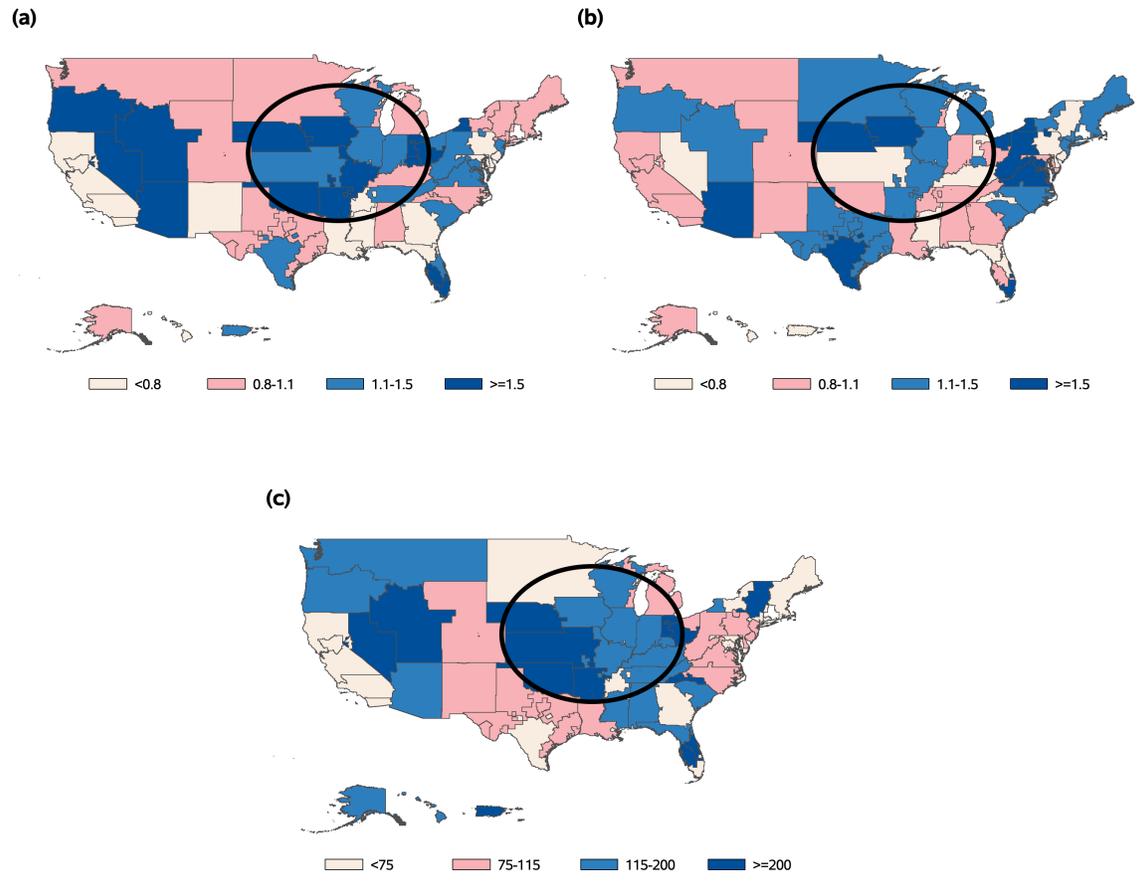}
    \caption{Maps of average Transplant Rate Ratio (TRR) posterior medians and the provider-level adjustment variable ($W_i$), by Donation Service Area (DSA). An example region with high donor organ availability is circled. (a) Average TRR posterior median among centers in the DSA, not adjusting for $W_i$, (b) Average TRR posterior median among centers in the DSA, adjusting for $W_i$, (c) $W_i$: Number of donors per 1000 transplant candidates within the DSA. }
    \label{fig:O_Map}
\end{figure}

The DSA-level map in Figure \ref{fig:O_Map}a shows that the transplant centers with the best evaluations, based on the SRTR's original posterior distribution, were concentrated in certain areas of the U.S. (e.g., the circled region in Figure \ref{fig:O_Map}). These regions also had some of the highest levels of donor organ availability in the nation (Figure \ref{fig:O_Map}c). After adjusting for $W_i$, this spatial correlation became less apparent (Figures \ref{fig:O_Map}b,c), providing evidence that our adjustment appropriately controls for the geographic confounding effects. 

\section{Discussion}

We have developed a privacy-preserving framework to estimate and adjust for the confounding effects of both observable and unobservable cluster-level risk factors when identifying cluster units with significant treatment effects. The proposed confounding effect estimator is highly robust to outlying cluster unit effects and only depends on summary information that is widely-used in the transplantation field. The proposed Pseudo-Bayesian inference procedure can incorporate these estimates while recognizing their statistical uncertainty and correcting for the inevitable influence of unobserved confounding. 

Many privacy-preserving models rely on distributed optimization algorithms, where each organization repeatedly calculates a set of statistics through multiple communications \citep{Jochems,Duan}. In this paper, we have shown that one may estimate cluster-level confounding effects simply by downloading publicly-available summary statistics from the SRTR and performing a single model-fitting routine. This provides a convenient alternative for stakeholders to refine provider evaluations, without any organization needing to share or reanalyze patient-level data. Even in applications where the full patient-level data are available, the proposed RPP-CLC model may be much more computationally efficient than robust CRE models, as shown in Section \ref{sec:cre_sim}.

Through our real data analyses, we have found that adjusting for geographic disparities in donor organ availability and other unobservable confounding factors, using our proposed methods, can substantially change the interpretations of U.S. transplant center performance. These new analyses are especially useful for efforts to expand access to transplantation and to find the sources of inequitable care. With our novel robust privacy-preserving model and inference procedure, researchers can conveniently and efficiently perform accurate assessments to improve the quality and equity of transplant care. 

\section*{Acknowledgements}

Research reported in this publication was partly supported by the National Institute of Diabetes and Digestive and Kidney Diseases under award number R01DK129539. This content is solely the responsibility of the authors and does not necessarily represent the official views of the National Institutes of Health. 

\bibliographystyle{apalike}
\bibliography{main}

\end{document}


\maketitle
\captionsetup[figure]{labelfont={bf},name={Web Figure},labelsep=period}
\captionsetup[table]{labelfont={bf},name={Web Table},labelsep=period}

\clearpage

\section*{\normalsize Web Appendix A: Derivation of the Functional Form for the Normal Distribution}
\label{sec:normal_appendix}

Assume that all expectations are conditional on $\sum_{j=1}^{n_i} b'(\theta^0_{ij})$, $\tilde{n}_i=\sum_{j=1}^{n_i} b''(\theta^0_{ij})$, and $\boldsymbol{W}_i$. Then, when $Y_{ij}$ follows a normal distribution,
\begin{align*}
E[Z_{FE,i}|\gamma^*_i=0]&=E[E[Z_{FE,i}|\alpha_i,\gamma^*_i=0]|\gamma^*_i=0] \nonumber \\
&=\frac{1}{\sqrt{n_i\sigma^2_{\varepsilon}}}E\left[\sum_{j=1}^{n_i} b'(\theta_{ij}^*)-\sum_{j=1}^{n_i} b'(\theta_{ij}^0)\bigg|\gamma_i^*=0\right] \\
&=\frac{1}{\sqrt{n_i\sigma^2_{\varepsilon}}}E\left[\sum_{j=1}^{n_i} \theta_{ij}^*-\sum_{j=1}^{n_i} \theta_{ij}^0\bigg|\gamma_i^*=0\right] \\
&=\sqrt{\frac{n_i}{\sigma^2_{\varepsilon}}}\boldsymbol{W}_i^{\top}\boldsymbol{\nu} \hspace{25pt} \text{(if $E[\alpha_i|\boldsymbol{W}_i,\boldsymbol{X}_{i1},\dots,\boldsymbol{X}_{in_i}]=0$)}.
\end{align*} 

\noindent Similarly, \begin{align*}
Var[Z_{FE,i}|\gamma^*_i=0]&=Var[E[Z_{FE,i}|\alpha_i,\gamma^*_i=0]|\gamma^*_i=0]+E[Var[Z_{FE,i}|\alpha_i,\gamma^*_i=0]|\gamma^*_i=0] \\
&=\frac{1}{n_i \sigma^2_{\varepsilon}} Var[n_i(\boldsymbol{W}_i^{\top}\boldsymbol{\nu}+\alpha_i)|\gamma_i^*=0]+\frac{1}{n_i\sigma_{\varepsilon}^2}E\left[Var\left[\sum_{j=1}^{n_i}Y_{ij}\right]\bigg|\gamma_i^*=0\right] \\
&=1+\frac{\sigma^2_{\alpha}}{\sigma^2_{\varepsilon}} n_i.
\end{align*}

\section*{\normalsize Web Appendix B: Derivation of the Functional Form for the Poisson Distribution}
\label{sec:poisson_appendix}

Under the Poisson model, $b(\cdot)=\exp(\cdot)$ and $a(\psi)=1$. Since $\alpha_i \sim \textrm{N}(0,\sigma_{\alpha}^2$), $\exp(\alpha_i) \sim \textrm{Lognormal}(0,\sigma^2_{\alpha}$). The null mean and variance functions follow directly from this result, combined with the approach in Web Appendix A. These arguments also apply to the Quasi-Poisson model, with $a(\psi)=\psi$, where $\psi$ may be different from one. 

\section*{\normalsize Web Appendix C: Derivation of the Approximate Functional Form for Exponential Family Distributions}
\label{sec:fam_appendix}

$$
E[Z_{FE,i}|\gamma^*_i=0]=\frac{1}{\sqrt{\tilde{n}_ia(\psi)}}E\left[\sum_{j=1}^{n_i} b'(\theta_{ij}^*)-\sum_{j=1}^{n_i} b'(\theta_{ij}^0)\bigg | \gamma_i^*=0\right]. $$

\noindent Applying the first-order Taylor series expansions in \citet{Hartman} (i.e. $b'(\theta^*_{ij})$ and $b''(\theta^*_{ij})$ around $\theta^0_{ij}$),\begin{align*}
E[Z_{FE,i}|\gamma^*_i=0] &\approx \frac{1}{\sqrt{\tilde{n}_ia(\psi)}} E\left[\tilde{n}_i(\theta^*_{ij}-\theta^0_{ij}) |\gamma_i^*=0\right] \\
&=\sqrt{\frac{\tilde{n}_i}{a(\psi)}}E[\boldsymbol{W}_i^{\top}\boldsymbol{\nu}+\alpha_i] \\
&=\sqrt{\frac{\tilde{n}_i}{a(\psi)}}\boldsymbol{W}_i^{\top} \boldsymbol{\nu} \hspace{25pt} \text{(if $E[\alpha_i|\boldsymbol{W}_i,\boldsymbol{X}_{i1},\dots,\boldsymbol{X}_{in_i}]=0$).} \\
\end{align*}

\noindent If $\gamma^*_i$ is random and $E[\gamma^*_i|\boldsymbol{W}_i,\boldsymbol{X}_{i1},\dots,\boldsymbol{X}_{in_i}]=0$, then we can replace $\alpha_i$ above with $\gamma_i=\alpha_i+\gamma^*_i$, and the result still holds. $Var[Z_{FE,i}|\gamma^*_i=0]$ is derived in a similar manner as above, using the variance decomposition and Taylor series approximation in \citet{Hartman}. For any outcome distribution in the exponential family, \begin{align*}
Var[Z_{FE,i}|\gamma^*_i=0]&\approx \frac{1}{\tilde{n}_ia(\psi)}Var[\tilde{n}_i(\boldsymbol{W}_i^{\top}\boldsymbol{\nu}+\alpha_i)] \\
&+ E\left[1+\frac{\sum_{j=1}^{n_i}b'''(\theta^0_{ij})}{\tilde{n}_i}(\boldsymbol{W}_i^{\top}\boldsymbol{\nu}+\alpha_i )\right] \\
&=1+\frac{\sum_{j=1}^{n_i}b'''(\theta^0_{ij})}{\tilde{n}_i} \boldsymbol{W}_i^{\top} \boldsymbol{\nu}+\varphi \tilde{n}_i,
\end{align*}

\noindent where $\varphi=\sigma^2_{\alpha}/a(\psi)$.

\section*{\normalsize Web Appendix D: Model Estimation Algorithm}
\label{sec:model_appendix}

\noindent Our proposed model fitting algorithm is summarized below on the next page (extended from \citet{Hartman} and written for Poisson outcomes):

\begin{algorithm}
\caption*{\textbf{Model Fitting Algorithm}}
\begin{algorithmic}[1]
\item Fit an initial robust linear regression model using M-estimation, with $Z_{FE,i}$ as the outcome and $\sqrt{\frac{\tilde{n}_i}{a(\psi)}}\boldsymbol{W}_i$ as the predictor vector \citep{Huber,Venables}. 
\item Obtain $\boldsymbol{\widehat{\nu}}^{(0)}$ and $\widehat{s}^{(0)}$ from the previous step, where $\widehat{s}^{(0)}$ is the estimated scale parameter. 
\item Let $\widehat{\sigma}^{2^{(0)}}_{\alpha}=\{(\widehat{s}^{(0)})^2-1-\textrm{median}(\boldsymbol{W}_1^{\top}\boldsymbol{\widehat{\nu}}^{(0)},\dots,\boldsymbol{W}_I^{\top}\boldsymbol{\widehat{\nu}}^{(0)})\}/\textrm{median}(\tilde{n}_1,\dots,\tilde{n}_I)$. 
\item Plug $\boldsymbol{\widehat{\nu}}^{(0)}$ and $\widehat{\sigma}^{2^{(0)}}_{\alpha}$ into the mean and variance functions in Section 2.2.2-2.2.4, and calculate 
$$A_i=\widehat{E}[Z_{FE,i}|\gamma_i^*=0]-1.96\sqrt{\widehat{Var}[Z_{FE,i}|\gamma^*_i=0]},$$
$$B_i=\widehat{E}[Z_{FE,i}|\gamma_i^*=0]+1.96\sqrt{\widehat{Var}[Z_{FE,i}|\gamma^*_i=0]}.$$
\item Generate a sequence of $\pi_0$ values from zero to one, denoted as the $1 \times N_{seq}$ vector $\boldsymbol{\pi}_0$. 
\For{$\iota=1,\dots,N_{seq}$} \State Maximize the log-likelihood, $\log\{L(\boldsymbol{\nu},\sigma^2_{\alpha}; \pi_0=\boldsymbol{\pi}_{0\iota})\}$, with respect to $\boldsymbol{\nu}$ and $\sigma^2_{\alpha}$, using the Nelder-Mead algorithm and $(\boldsymbol{\widehat{\nu}}^{(0)},\widehat{\sigma}^{2^{(0)}}_{\alpha})$ as the starting values \citep{NM}. Obtain $(\boldsymbol{\widehat{\nu}},\widehat{\sigma}^2_{\alpha})_{\iota}=\underset{\boldsymbol{\nu},\sigma^2_{\alpha}}{argmax}(\log\{L(\boldsymbol{\nu},\sigma^2_{\alpha}; \pi_0=\boldsymbol{\pi}_{0\iota})\})$.
\EndFor
\item Find $\iota^*=\underset{\iota}{argmax}(log\{L((\boldsymbol{\widehat{\nu}},\widehat{\sigma}^2_{\alpha})_{\iota}; \pi_0=\boldsymbol{\pi_{0\iota}})\})$. \\
$\widehat{\pi}_0=\boldsymbol{\pi}_{0\iota^*}$ and 
$(\boldsymbol{\widehat{\nu}},\widehat{\sigma}^2_{\alpha})=(\boldsymbol{\widehat{\nu}},\widehat{\sigma}^2_{\alpha})_{\iota^*}$.
\end{algorithmic}
\end{algorithm}

\clearpage

\section*{\normalsize Web Appendix E: Proof of the Connection With the CRE Model}
\label{sec:cre_appendix}

Under the CRE assumptions, $\gamma_i^*=\boldsymbol{\bar{X}}_i^{\top}\boldsymbol{\xi}+\tau_i$, where $\tau_i|\boldsymbol{X}_{i1},\dots,\boldsymbol{X}_{in_i} \sim N(0,\sigma^2_{\tau})$. Therefore, $E[\tau_i|\boldsymbol{X}_{i1},\dots,\boldsymbol{X}_{in_i}]=0$ and $\theta^*_{ij}-\theta^0_{ij}=\tau_i+\boldsymbol{\bar{X}}_i^{\top}\boldsymbol{\xi}$. It follows directly from the arguments of Web Appendix C (substituting $\tau_i$ for $\alpha_i$ and $\boldsymbol{\bar{X}}_i$ for $\boldsymbol{W}_i$) that $$E[Z_{FE,i}|\gamma^*_i=0]\approx \sqrt{\frac{\tilde{n}_i}{a(\psi)}}\boldsymbol{\bar{X}}_i^{\top}\boldsymbol{\xi}.$$

\noindent The exact mean and variance functions for the normal and Poisson models follow directly from the arguments above, combined with those from Web Appendices A and B. 

\section*{\normalsize Web Appendix F: Estimator for the Variance of the Confounding Effect Estimate}
\label{sec:bayesian_variance_appendix}

Let $\boldsymbol{W}$ be an $I$ by $P$ matrix of the observed center-level confounding variables, where $P$ is the number of observed variables, and let $\boldsymbol{W_{\textrm{null}}}$ be the submatrix that corresponds to null providers. Since we do not know exactly which providers are null, we obtain $\boldsymbol{W_{\textrm{null}}}$ by selecting all rows of $\boldsymbol{W}$ that correspond to $Z_{FE,i}$ values falling within the assumed null intervals $[A_i, B_i]$ from Section 2.2.5. From (5), $Z_{FE,i}$ approximately follows a linear model with heteroskedastic errors for all centers included in $\boldsymbol{W_{\textrm{null}}}$. For the Poisson model, the variance of the $i^{th}$ error term is given by $1+\boldsymbol{W}_i^{\top} \boldsymbol{\nu}+\varphi \tilde{n}_i$, since $b''(\theta^0_{ijk})=b'''(\theta^0_{ijk})=\exp(\theta^0_{ijk})$. Applying the covariance estimator formula from \citet{White}, $\boldsymbol{\Sigma_{\widehat{\nu}}}$ is consistently estimated by $$
  \boldsymbol{\widehat{\Sigma}_{\widehat{\nu}}}=(\boldsymbol{W_{\textrm{null}}}^{\top}\boldsymbol{W_{\textrm{null}}})^{-1}\boldsymbol{W_{\textrm{null}}}^{\top}\boldsymbol{\widehat{\Omega} W_{\textrm{null}}}(\boldsymbol{W_{\textrm{null}}}^{\top}\boldsymbol{W_{\textrm{null}}})^{-1},
$$

\noindent where $\boldsymbol{\widehat{\Omega}}_{ii}=1+\boldsymbol{W}_i^{\top} \boldsymbol{\nu}+\widehat{\varphi} \tilde{n}_i$ and $\boldsymbol{\widehat{\Omega}}_{i \ne j}=0$. 

\section*{\normalsize Web Appendix G: Derivation of the Pseudo-Bayesian Flagging Approach}
\label{sec:bayesian_appendix}

Under the SRTR's Gamma-Poisson model, it is assumed that $O_i|E_i \sim \textrm{Poisson}(R_i E_i)$, where $R_i \sim \textrm{Gamma}(2,2)$. This yields the SRTR's posterior distribution ($f_{R_i}(r_i)$), $$R_i | O_i, E_i \sim \textrm{Gamma}(O_i+2,E_i+2).$$

\noindent Under our proposed model, $O_i|E_i,\Lambda_i \sim \textrm{Poisson}(R_i^* E_i \lambda_i)$, where $R_i^* \sim \textrm{Gamma}(2,2)$. This yields the following conditional posterior distribution ($f_{R_i^*|\Lambda_i}(r_i^*|\lambda_i)$): $$R_i^* | O_i, E_i, \Lambda_i \sim \textrm{Gamma}(O_i+2,E_i\lambda_i+2).$$

\noindent After defining the distribution of $\Lambda_i|O_i,E_i$, which we denote as $f_{\Lambda_i(\lambda_i)}$, we apply the Law of Total Probability to derive the distribution of $R_i^*|O_i,E_i$: $$f_{R^*_i}(r^*_i)=\int f_{R_i^*|\Lambda_i}(r^*_i|\lambda_i)f_{\Lambda_i}(\lambda_i) d\lambda_i.$$

\begin{figure}
    \centering
    \includegraphics[width=\textwidth]{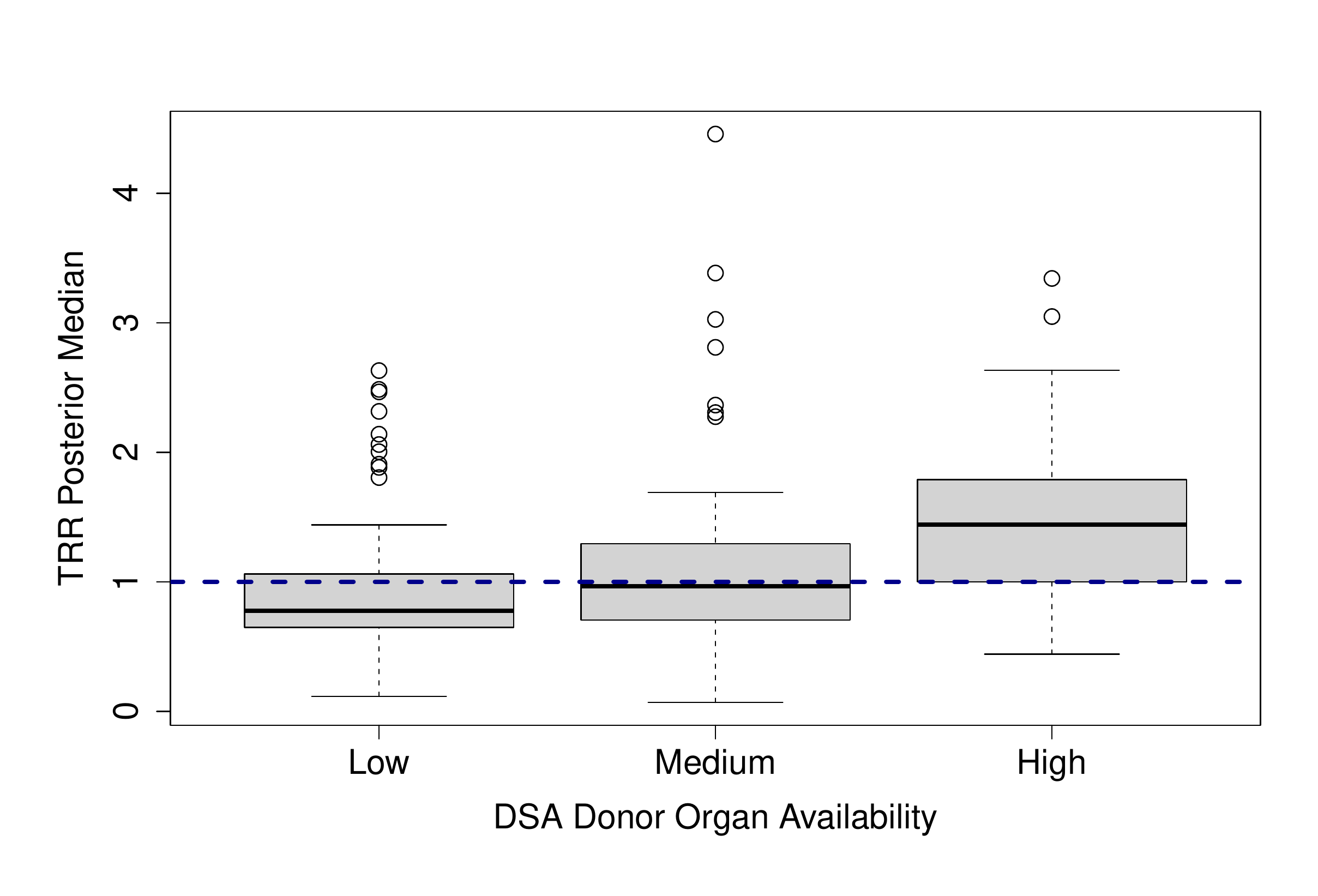}
    \caption{Distribution of the posterior medians for the Transplant Rate Ratio (TRR) measure, stratified by donor organ availability within the centers' Donation Service Areas (DSAs). The posterior median for each transplant center is computed from the SRTR's proposed Gamma distribution, which does not account for geographic variation in donor organ availability. Donor organ availability is defined as the number of donors per transplant candidate within a center's DSA (``Low," ``Medium," and ``High" groups are formed by tertiles). The horizontal dashed line is the reference line of TRR=1, which indicates that a center is performing consistently with national norms, and higher TRR values indicate better performance in terms of access to transplantation. Centers that reside in DSAs with high donor organ availability tend to have better evaluations based on the SRTR's metrics.}
    \label{fig:distributions}
\end{figure}

\bibliographystyle{apalike}
\bibliography{supp}